\documentclass[a4paper,11pt]{article}
\pdfoutput=1 

\usepackage{jheppub} 

\usepackage[T1]{fontenc} 

\usepackage{orcidlink}

\newcommand{\beq}{\begin{equation}}
\newcommand{\eeq}{\end{equation}}
\newcommand{\beqa}{\begin{eqnarray}}
\newcommand{\eeqa}{\end{eqnarray}}  
\newcommand{\beqar}{\begin{eqnarray*}}
\newcommand{\eeqar}{\end{eqnarray*}}

\newcommand{\ie}{{\em i.e.,}\ }
\newcommand{\mpT}{\vec p_{\rm T}^{\rm \,miss}}

\title{ Reconstructing a Heavy Neutral Lepton at the LHC}

\author[1]{Pablo de la Torre\orcidlink{0009-0008-4998-2304},\note{Corresponding author.}}
\author{Manuel Masip\orcidlink{0000-0002-7750-2514} and}
\author{Fuensanta Vilches\orcidlink{0009-0008-3245-0520}}

\affiliation{Departamento de F{\'\i}sica Te\'orica y del Cosmos,\\Universidad de Granada, E-18071 Granada, Spain}

\emailAdd{pdelatorre@ugr.es}
\emailAdd{masip@ugr.es}
\emailAdd{fuenvilches@ugr.es}

\abstract{
Heavy lepton singlets $N$ slightly mixed with a standard neutrino $\nu_\ell$  
are usually searched for at the LHC
in the trilepton plus $p_{\rm T}^{\rm miss}$ channel: 
$pp \to W^+ \to  \ell^+ N$ with $N\to \ell^- W^+ \to  \ell^- \ell'^+ \nu$. We
show that, although the longitudinal momentum of the final $\nu$ escapes detection, 
the mass of the heavy lepton can be reconstructed. While this possibility has not been
considered in recent LHC searches, we find that
the search for a mass peak could improve the current collider bounds on the mixing 
$|V_{\ell N}|^2$ for any mass $m_N\ge  M_W$.
}

\begin{document}
\maketitle
\flushbottom

\section{Introduction} 
Heavy neutral leptons (HNL) provide the simplest UV completion of the Standard Model (SM) that is 
able to generate the dimension 5 Weinberg operator \cite{Weinberg:1979sa}, which arguably makes them
the most motivated scenario for BSM physics \cite{Fukugita:1986hr,Dodelson:1993je,Atre:2009rg,Tello:2010am,Das:2012ze,Boyarsky:2018tvu,Drewes:2022akb}. Let us briefly discuss a minimal setup (see also the models in \cite{Cuesta:2021kca,delaTorre:2023nfk}) 
PMNSusing two-component spinors.

To explain neutrino masses we need at least two 
 bi-spinors  $(N, N^c)$ of opposite lepton number $L =\pm 1$ combined into a Dirac field of mass $M$. Assuming no extra Higgs bosons, lepton-number conservation  and only dim $\le 4$ operators we have
\beq
-{\cal L} \supset M \, N N^c + y_\nu\, H L N^c + {\rm h.c.}\,,
\eeq
where $H=(h^+\; h^0)$, $L=(\nu \; \ell)$ and we have redefined the three lepton doublets to obtain the flavor combination in the Yukawa interaction. After electroweak (EW) symmetry breaking the neutrino mass matrix reads
\beq
{\cal M} = \begin{pmatrix}
          &  &   & \cdot   & 0  \\
          & 0  &  & \cdot  & 0  \\
         &  &  & \cdot  & m \\
        \cdot   & \cdot  & \cdot  & \cdot  & M   \\
        0 & 0 & m & M & \cdot 
\end{pmatrix},
\label{nu1}
\eeq
where $m=y_\nu v/\sqrt{2}$ and the dots indicate terms forbidden by lepton number conservation. This rank-2 matrix implies a Dirac field $\, (N', N^c)\, \,$,  with $\, N'=c_\alpha N + s_\alpha \nu_3 \,\,$,  of mass $m_{N}=\sqrt{M^2+m^2}$ plus three massless neutrinos. The  neutrino $\nu'_3=-s_\alpha N + c_\alpha \nu_3$ is massless but has now a small component $s_\alpha= m/\sqrt{M^2+m^2}$ along the
sterile flavor $N$. 
 
In order to get non-zero masses we need to break lepton number conservation: we can add small Majorana masses 
($\mu_{1,2}\ll M$, $\Delta L =2$) for
the two heavy modes and suppressed Yukawas  ($\tilde y_\nu \ll y_\nu$, $\Delta L =1$) for $N$,
implying 
\beq
{\cal M} = \begin{pmatrix}
          &  &   & 0   & 0  \\
          & 0  &  & \mu_3 & 0  \\
         &  &  & \mu'_3  & m \\
       0  & \mu_3  & \mu'_3  & \mu_1  & M   \\
        0 & 0 & m & M & \mu_2
\end{pmatrix}.
\label{nu2}
\eeq
The mass 
$\mu_1$ increases in one unit the rank of ${\cal M}$ and defines an inverse seesaw with
$m_{\nu_3}\approx \mu_1 (m/M)^2$ \cite{Mohapatra:1986bd}.
The term $\mu_2$ does not give mass to a second standard 
neutrino, it just breaks the degeneracy between the two (now Majorana) heavy
neutrinos. The term $\mu_3=\tilde y_\nu v/\sqrt{2}$ is then necessary to give a mass 
$m_{\nu_2}\approx \mu_3^2/\mu_1$ to $\nu'_2$, with 
$\nu'_1$ staying massless. 
Alternatively, $\mu_{3}$ could be forbidden by a discrete symmetry  and $m_{\nu_2}$ obtained by adding a second heavy neutrino pair ({\it i.e.}, an inverse seesaw for the two massive neutrino families, with the third neutrino massless).

The mass parameters $M$ and $\mu_{1,2}$ above are not proportional to the EW scale, and the matrix in Eq.~(\ref{nu2}) may accommodate any values between $10^{-3}$ and $10^{15}$ GeV: any HNL mass seems equally {\it natural}, and only the data may have a say about Nature's choice. 
Masses larger than TeV, however, imply a very small mixing $s_\alpha$ and thus a decoupled HNL. In particular, the usual seesaw mechanism is obtained if the three mass parameters and the Yukawa couplings are all unsuppressed, {\it e.g.}, 
$M, \mu_{1,2} \approx 10^{10}$ GeV and $m,\mu_3 \approx 1$ GeV. 

\sloppy
The possibility of two Majorana HNLs at the TeV scale with sizeable heavy-light mixings ({\it e.g.}, $m=10$ GeV, 
$\mu_{1,2}\approx 1$ TeV and $M=0$) requires a fine tuning that is not stable under radiative corrections ($m^2/\mu_2+\mu'^2_3/\mu_1\approx 10^{-8} m^2/\mu_2$). If one of the HNLs is significantly heavier than the other one,
the cancelation forces its mixing to be smaller and the model reduces to a single Majorana HNL mixed with 
a combination of the standard neutrinos. Throughout our analysis we will assume a quasi-Dirac HNL $N$.

The heavy-light mixing $s_\alpha$ defines then the couplings of $\nu_\ell$ and $N$ 
to the $W^\pm$ and $Z$ bosons (we drop the prime to denote mass eigenstates). In particular, $\nu_\ell$ will see its gauge couplings reduced by a factor of $c_\alpha$ whereas $N$ will now couple to the $W$ with a strength proportional to $V_{\ell N} \approx s_\alpha$. The model also implies heavy--light couplings both to the $Z$ boson ($\propto s_\alpha c_\alpha$) and
to the Higgs boson ($y_\nu=\sqrt{2} s_\alpha M/v$). We will consider the case where $N$ couples to a single generation of SM neutrinos  \cite{Beacham:2019nyx, Drewes:2022akb}. 

Collider bounds on $|V_{\ell N}|^2$ may be obtained at energies below or above $m_N$. At lower energies the HNL is not produced and bounds arise from observables like weak interaction universality, precision observables related to muon decays (the mixing with $\nu_\mu$ changes the definition of $G_F$), the invisible
width of the $Z$ boson or one-loop flavor changing processes like $\mu \to e \gamma$. A global fit of these observables, which probe deviations from the unitarity of the Pontecorvo-Maki-Nakagawa-Sakata (PMNS) matrix due to heavy neutrino mixing, sets limits on $|V_{\ell N}|^2$ ranging from $10^{-2}$ to $10^{-4}$, with the strongest bounds for $\ell=\mu$ \cite{Fernandez-Martinez:2016lgt,Hernandez-Tome:2018fbq,Hernandez-Tome:2019lkb, Blennow:2023mqx}. 

\begin{figure}[!t]
\begin{center}
\includegraphics[scale=0.9]{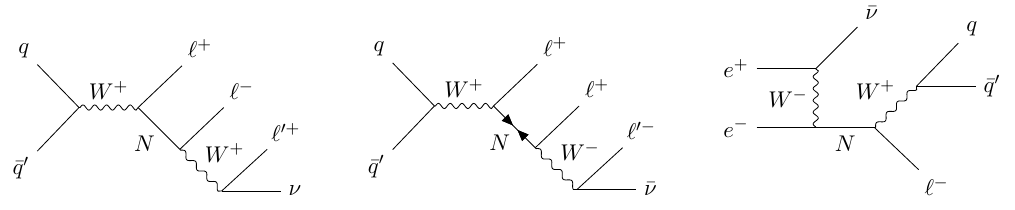}
\end{center}
\vspace{-0.5cm}
\caption{Dominant diagrams in $N$ searches at LHC in the Dirac (left) and Majorana (center) cases and 
at LEP (right).
\label{f1}}
\end{figure}
Here we will discuss higher energy processes with the direct production of the HNL. The signatures of such models at colliders have been studied extensively \cite{delAguila:2008cj, Deppisch:2015qwa, Cai:2017mow, Das:2017nvm, Das:2017gke, Bhardwaj:2018lma, Pascoli:2018heg, Abdullahi:2022jlv, Antel:2023hkf}. The current bounds on $|V_{\ell N}|^2$ for  $m_N$ above the $W$ mass have been obtained by CMS in the trilepton channel at the LHC \cite{CMS:2018iaf,CMS:2024xdq} and by DELPHI \cite{DELPHI:1996qcc} and L3 \cite{L3:2001zfe} at LEP 
(only for $\ell=e$). We will focus on the 
trilepton channel, a process dominated by the charged-current Drell--Yan
$q \bar{q}' \to W^{\pm} \to N l^{\pm}$ \cite{Keung:1983uu,  Petcov:1984nf}  (see Fig.~\ref{f1}). In our analysis we will 
also include vector boson fusion $q \gamma \to N l^{\pm} q'$ \cite{Datta:1993nm, Dev:2013wba, Alva:2014gxa}, a process giving a significant contribution for $m_N \geq 500$ GeV. We will show that, although the longitudinal momentum of the $\nu$ escapes detection, the mass $m_N$ of the HNL can be reconstructed.

\begin{figure}[!t]
\begin{center}
\includegraphics[scale=0.95]{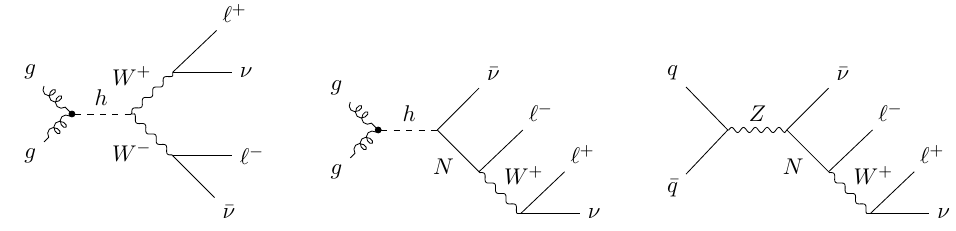}
\end{center}
\vspace{-0.5cm}
\caption{Higgs to $W^+ W^-$ in the SM and possible HNL contributions to dilepton plus $p^{\rm miss}_{\rm T}$.
\label{f2}}
\end{figure}

We will comment as well on the possible effects introduced by an HNL 
 in Higgs searches at the LHC. In particular, 
notice (see Fig.~\ref{f2}) that the dilepton plus missing $p_{\rm T}$ signal from $h\to W W^*$ observed at CMS 
coincides with the one from $h,Z\to \nu N$, although 
the kinematics in each process is obviously different. 

\section{HNL at the LHC} 
To estimate the possible signal of an HNL at the LHC, we first produce a next-to-leading-order (NLO) UFO file \cite{Degrande:2011ua} with the new vertices using the \textsc{Mathematica} packages \textsc{FeynRules} \cite{Alloul:2013bka}, \textsc{NLOCT} \cite{Degrande:2014vpa} and \textsc{FeynArts} \cite{Hahn:2000kx}.
 Then, we generate signal and background events at NLO using \textsc{MadGraph5\_aMC@NLO} v2.9.7 \cite{Alwall:2014hca} interfaced with \textsc{Pythia} v8.310 \cite{Bierlich:2022pfr}, \textsc{MadSpin} \cite{Artoisenet:2012st} and \textsc{Delphes} v3.2.0 \cite{deFavereau:2013fsa}. In \textsc{Delphes} we have used the CMS simulation card, including a photon conversion module based on the characteristics of the detector 
\cite{Veszpremi:2014hpa, CMS:2015myp, Brenner, Sirunyan_2021}. For the parton distribution functions (PDFs), we use  NNPDF3.1 \cite{NNPDF:2017mvq} in all the cases except for vector boson fusion signal process, where we use NNPDF31.luxQED \cite{Manohar:2016nzj, Manohar:2017eqh, Bertone:2017bme} (both as implemented in \textsc{lhapdf} \cite{Buckley:2014ana}). For the merging between jets from matrix elements and from parton showers, we have used the FxFx matching scheme \cite{Frederix:2012ps}. Finally, we have analyzed the results
with custom routines based on FasJet \cite{Cacciari:2011ma}, ROOT \cite{Brun:1997pa} and HepMC \cite{Buckley:2019xhk}. 

After we generate the events, we apply the cuts defined in \cite{CMS:2024xdq} for the high-mass regime, with at least one opposite-sign same-flavour (OSSF) lepton pair and

\begin{enumerate}
    \item The three leptons must be isolated as defined in \cite{Rehermann:2010vq}, with
$\, \, p_{\rm T}(\ell_1)>55 $ GeV, $ \, \, \, \, p_{\rm T}(\ell_2)>15 \,$ GeV, $p_{\rm T}(\ell_3)>10$ GeV.
    \item No $b$ jets.
    \item Any OSSF pair must satisfy $m(\ell^+ \ell^-)>5$ GeV and $\left| m(\ell^+ \ell^-)-m_Z \right| >15$ GeV.
    \item The trilepton invariant mass is $\left| m(3\ell)-m_Z \right| >15$ GeV.
\end{enumerate}

\begin{table}[t!]
\begin{center}
\begin{tabular}{|c|ccccc|}
\hline
 $m_N$ [GeV] & 90  & 130  & 170 & 400 & 600  \\ 
\hline
$\mu^\pm \mu^\mp e^\pm$ events & 858 &  297 & 101  & 4.01 & 1.78 \\
\hline
events after cuts  & 148 & 119 & 72.5 & 3.87 &  1.73 \\
 & (17.2\%) & (40.0\%) & (71.8\%) & (96.5\%) & (97.2\%) \\
\hline
\end{tabular} 
\caption{Signal events in the channel $\mu^\pm \mu^\mp e^\pm\nu$ before and after cuts (138 fb$^{-1}$ at 13 TeV) for several HNL masses and $|V_{\mu N}|^2=10^{-2}$.}
\label{Cuts1}
\end{center}
\end{table}

To illustrate our results, let us consider three benchmarks masses $m_N$ in the  $90$--$170$ GeV range and two more in the higher mass region $400$--$600$ GeV. In all the cases we assume a mixing $|V_{\mu N}|^2=10^{-2}$ with the muon flavor
(the cross sections are proportional to $|V_{\mu N}|^2$) and  we take a luminosity of 138 fb$^{-1}$ at 13 TeV. 
We first focus on the channel 
\beq
pp \to \mu^\pm N \to \mu^\pm \mu^\mp W^\pm \to \mu^\pm \mu^\mp e^\pm \nu \,.
\eeq
In our simulation we include events where the final electron comes from the leptonic decay of a tau lepton (\ie the $\ell'^+$ in Fig.~\ref{f1} may be an $e^+$ or a $\tau^+$ decaying into  $\bar \nu_\tau \,e^+ \nu_e$). Notice that in this 
quasi-Dirac case, a lepton chain with the first two leptons having the same charge is not allowed (see Fig.~\ref{f1}). Table \ref{Cuts1} lists our estimates for the total number of events in this channel before and after cuts, which agree within a 5\% with the ones expected by CMS in \cite{CMS:2024xdq} for the same cuts and luminosity (see Table {\it Predicted signal yields (High mass region Hb, 
$\mu$-coupling)} in the HEPData record of the CMS analysis \cite{CMSrepository}).
 The relatively low efficiency of the cuts at $m_N=90$--$130$ GeV suggests the analysis has been optimized for higher HNL masses\footnote{For $m_N=90$ GeV, in particular, we find that  a reduction from 10 to 5 GeV in the minimum $p_T$ of the least energetic lepton increases the signal by a factor of 1.42 and the background by a factor of 1.12.}. 

 \begin{figure}[t!]
\begin{center}
\includegraphics[scale=0.68]{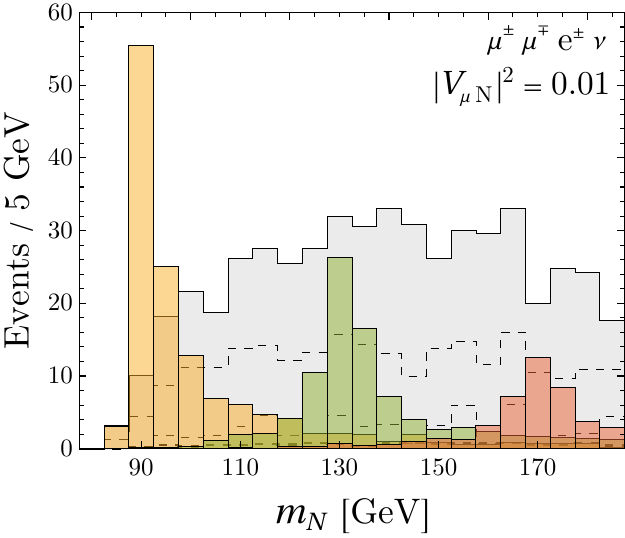}\hspace{0.5cm}
\includegraphics[scale=0.68]{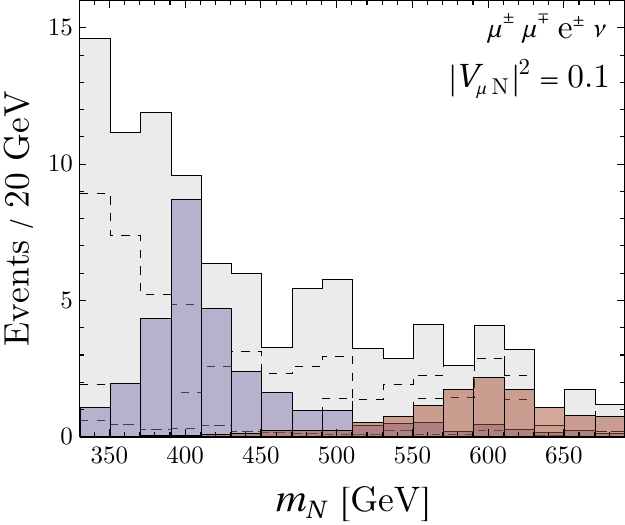}
\end{center}
\vspace{-0.5cm}
\caption{Reconstruction of $m_N$ from $pp \to \mu^\pm \mu^\mp e^\pm \nu$ in the low (left) and high (right) mass regions for $|V_{\mu N}|^2=0.01$ and $0.1$, respectively, together with the stacked backgrounds (from top to bottom, the  contributions from $Z\gamma$, $WZ$, $ZZ$ and nonprompt+other separated by dashes).
\label{f3}}
\vspace{-0.3cm} 
\end{figure}

The background, in turn, is made up of $WZ$ (37\%), $ZZ$ (1\%), $Z\gamma$ (38\%), nonprompt ($Z$+jets, $t \bar{t}$+jets, 20\%) and other ($WWW$, $WWZ$, $ZZ$, $t\bar{t}W$, $t\bar{t}Z$, $t\bar{t}\gamma$, 4\%). We obtain 
 over 54900 events that are reduced to 1260 background events after the cuts.
These numbers are consistent (within a 20\%, see Table {\it Data and background yields} in \cite{CMSrepository}) with the ones used by CMS in \cite{CMS:2024xdq}. 

Now, the detector provides the momenta
of the two muons and of the electron in all the signal events, 
whereas the transverse momentum $\vec p^{\, \nu}_{\rm T}$ of the neutrino can be obtained from $\mpT$. In addition, 
we notice that for $m_N>m_W$ the $W$ boson decaying into electron plus neutrino is on shell. This can be used to deduce the longitudinal momentum $p_{\rm L}^\nu$ of the neutrino (see \cite{Sanyal:2023pfs} for an analogous calculation in a different context): it is  solution to the quadratic equation
\beq
(p^e_{\rm T})^2 \, (p_{\rm L}^\nu)^2 - p^e_{\rm L} \left( m_W^2 + 2 \, \vec p^{\,e}_{\rm T} \cdot \vec p^{\,\nu}_{\rm T} \right)  p_{\rm L}^\nu - \left( {m_W^4\over 4} 
+ m_W^2 \, \vec p^{\,e}_{\rm T} \cdot \vec p^{\,\nu}_{\rm T}+ (\vec p^{\,e}_{\rm T} \cdot \vec p^{\,\nu}_{\rm T})^2 - (p^e)^2\, (p^\nu_{\rm T})^2 \right) =0\,.
\label{pT}
\eeq
With the complete momenta of the neutrino and of the electron and the muon of opposite charge  (i.e., $e^\pm \mu^\mp$) we can now reconstruct $m_N$,
\beq
m_N = \sqrt{\left( E^\mu+E^e+E^\nu \right)^2 - 
\left( \vec p^{\,\mu} + \vec p^{\,e} + \vec p^{\,\nu} \right)^2}\,.
\eeq
We find, however, that after the detector simulation 8\% of the signal events for $m_N=90$ GeV or 13\% for 600 GeV have no real solution to Eq~(\ref{pT}). This introduces an effective extra cut in the signal that, interestingly, is stronger on the background: 25\% of the background events are cut by the requirement of having a solution to Eq~(\ref{pT}). 
For the rest of events the equation implies a two-fold degeneracy, and we find that the lower solution provides the correct value of $p_{\rm L}^\nu$ 87\% of the times.

\begin{table}[b!]
\begin{center}
\begin{tabular}{|c|ccccc|}
\hline
 $m_N$ [GeV] & 90  & 130  & 170 & 400 & 600  \\ 
\hline
$\mu^\pm \mu^\mp \mu^\pm$ events & 981 &  336 & 112  & 4.39 & 1.90 \\
\hline
events after cuts  & 164 & 92.3 & 53.7 & 4.13 &  1.85 \\
 & (16.7\%) & (27.4\%) & (47.9\%) & (94.1\%) & (97.3\%) \\
\hline
\end{tabular} 
\caption{Signal events in the channel $\mu^\pm \mu^\mp \mu^\pm\nu$ before and after cuts (138 fb$^{-1}$ at 13 TeV) for several HNL masses and $|V_{\mu N}|^2=10^{-2}$.}
\label{Cuts2}
\end{center}
\end{table}

In Fig.~\ref{f3} we plot the reconstruction of different values of $m_N$ in the low and high mass regions together with the background.

We can increase the data sample if we include the possibility that the $W$ boson decays into a muon, 
\beq
pp \to \mu^\pm N \to \mu^\pm \mu^\mp W^\pm \to \mu^\pm \mu^\mp \mu^\pm \nu \,.
\eeq

We plot the number of events before and after cuts in Table \ref{Cuts2}. This trimuon channel has an estimated background of 31320 events (60\% from $WZ$, 36\% from $ZZ$, 3\% nonprompt and below 1\% from other)
that is reduced to 640 events after cuts.

In this case, the muon needed to reconstruct $m_N$ appears together with a second muon with the same charge. We notice, however, that the production of
an HNL (see Fig.~\ref{f1}) is favored by the collision of a valence quark and a sea antiquark in the initial protons. This 
tends to give a larger $p_{\rm T}$ to the muon in the primary vertex and implies that $m_N$ must be reconstructed with the muon of lower $p_{\rm T}$. We find that this prescription works 92\% of the times for 
$m_N=130$ GeV or 77\% for $m_N=400$ GeV.

\begin{figure}[t!]
\begin{center}
\includegraphics[scale=0.68]{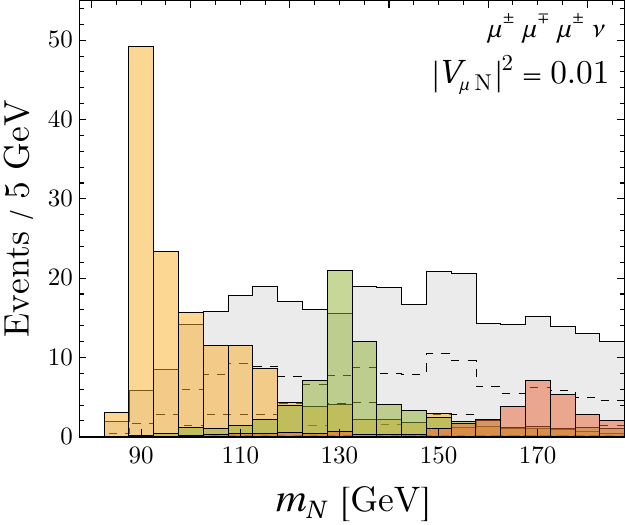}\hspace{0.5cm}
\includegraphics[scale=0.68]{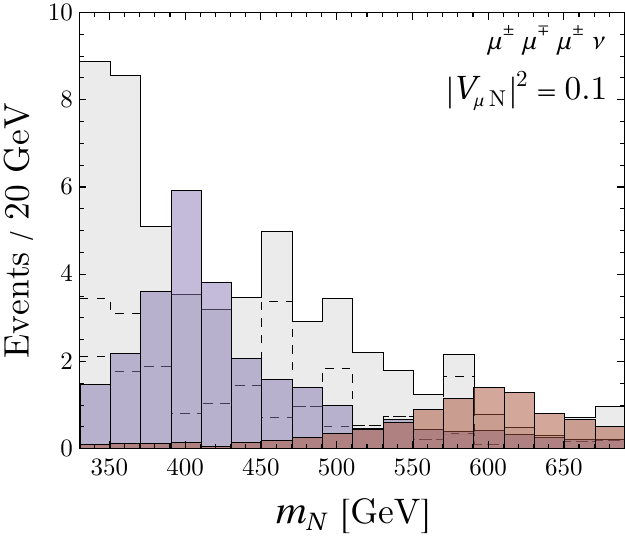}
\end{center}
\vspace{-0.5cm}
\caption{Reconstruction of $m_N$ from $pp \to \mu^\pm \mu^\mp \mu^\pm \nu$ in the low (left) and high (right) mass regimes for $|V_{\mu N}|^2=0.01$ and $0.1$, respectively,  together with the stacked backgrounds (from top to bottom, the  contributions from $WZ$, $ZZ$ and nonprompt+other separated by dashes).
\label{f7}}
\end{figure}

Therefore we consider as well this trimuon channel. The reconstruction is done with the same criteria as before but changing the electron by the muon with less $p_{\rm T}$ among the two with same charge. The  distribution for the reconstructed $m_N$ is shown in Fig.~\ref{f7} in the two mass regimes. Combining the $\mu\mu e$ and the $\mu\mu\mu$ channels, we obtain an estimate of the expected LHC limits at 95\% confidence level (we use the $CL_s$ prescription \cite{Read:2002hq} and assume that the impact of systematics on the total uncertainty is 35\% \cite{CMS:2024xdq}). We find bounds (see Table \ref{bounds}) that go from 
$|V_{\mu N}|^2<8.0\times 10^{-4}$ for $m_N=90$ GeV to 
$|V_{\mu N}|^2<6.3\times 10^{-2}$ for $m_N=600$ GeV.
To illustrate how robust these results are, we have repeated the analysis introducing a $10\%$ additional smearing of unknown origin (pile-up, detector noise, imperfect jet energy resolution and scale) in the reconstruction of the
missing transverse momentum: a random 10\% variation\footnote{Distributed according to a Gaussian of 
$\sigma=0.1\,p_{T}^{\rm miss}$.} in each component $ \left( p^{\rm miss}_{x,y} \right)$ of the transverse momentum. We find that for $m_N=130$ GeV 
the expected bound changes a 22\%, from 
$|V_{\mu N}|^2<2.6\times 10^{-3}$ to $|V_{\mu N}|^2<3.2\times 10^{-3}$. This weaker bound is mainly caused by a reduction from $90\%$ to $80\%$ in the fraction of signal events that provide a solution to $p_L^\nu$ in Eq.~(\ref{pT}).

We can estimate the possible relevance of this observable in current collider searches.
In a recent analysis \cite{CMS:2024xdq}, CMS set expected bounds using boosted decision trees (BDT) trained with several kinematic observables ($\Delta \text{R}\left[\text{min }m(l^+l^-)\right]$, $m_T$, $p_T(l_3)$ and 
$p_{T}^{\text{miss}}$). 
Our results indicate that this optimized combination of observables provides a less efficient signal-background discrimination than $m_N$, especially in the low mass region: they expect (see Table {\it Limits on Dirac HNL with muon coupling} in the HEPData record of the CMS analysis \cite{CMSrepository}) a  
limit $|V_{\mu N}|^2<2.2\times 10^{-3}$ (versus $8.0\times 10^{-4}$) for a mass $m_N=90$ GeV or $|V_{\mu N}|^2<5.6\times 10^{-3}$ (versus $2.6\times 10^{-3}$) for $m_N=130$ GeV. Therefore, significantly better bounds could be expected if the reconstructed $m_N$ were added to the pool of observables training the BDT.

\begin{table}[t!]
\begin{center}
\begin{tabular}{|c|ccccc|}
\hline
 $m_N$ [GeV] & 90  & 130  & 170 & 400 & 600  \\ 
\hline
$|V_{\mu N}|^2$ limit & $8.0\times 10^{-4}$ 
&  $2.6\times 10^{-3}$ & $5.0\times 10^{-3}$  & $3.9\times 10^{-2}$ & $6.3\times 10^{-2}$  \\
\hline
\end{tabular} 
\caption{Expected bounds on $|V_{\mu N}|^2$ by combining the channels $\mu^\pm \mu^\mp e^\pm\nu$ and 
$\mu^\pm \mu^\mp \mu^\pm\nu$ for 138 fb$^{-1}$ at 13 TeV.}
\label{bounds}
\end{center}
\end{table}

An analogous analysis can be done for  the mixing $|V_{e N}|$,
\beq
pp \to W^\pm\to e^\pm N \to e^\pm e^\mp \,W^\pm \to e^\pm e^\mp \ell^\pm \, \nu \,.
\eeq
Our results suggest similar bounds in this case. For example, we obtain a  20\% smaller signal after cuts at $m_N=130$ GeV within a 10\% smaller  background, implying an expected bound around
 $|V_{e N}|^2<2.9\times 10^{-3}$ for this HNL mass.

\section{Other production channels} 

Finally, we would like to comment on the possible reconstruction of the HNL mass also in other channels. In particular, we would like to argue that in Higgs searches $h\to W W^* \to \mu^+ \nu \,e^- \bar \nu$ (dilepton of different flavor plus missing $p_{\rm T}$) 
\cite{ATLAS:2018xbv,ATLAS:2019vrd,CMS:2020dvg,ATLAS:2021pkb,ATLAS:2023hyd} 
with slightly different cuts could be sensitive to the presence of a HNL produced through Higgs and $Z$ boson in the 
$s$ channel (see Fig.~\ref{f2}): 
\beq
pp \to Z(h) \to \nu N\to \nu\,\mu^+ W^-  \to \nu\, \mu^+ \bar \nu\, e^-.
\eeq

Take $m_N=85$ GeV (the argument applies to any masses between $m_W$ and around $100$ GeV) 
and $|V_{\mu N}|^2=10^{-3}$. Before the cuts described in \cite{CMS:2022uhn}, we estimate 9950 HNL events versus 11320 Higgs events. 
 We notice that 
the low value of $m_N$  favors that the s-channel $Z$ boson in Fig.~\ref{f2} is near the mass shell, implying that the neutrino
from $Z\to \nu N$ will carry little $p_{\rm T}$. We can then assume that most of the missing $p_{\rm T}$ is carried by the second neutrino and, as before, use that the parent $W$ is on shell to find $p^\nu_{L}$ and reconstruct $m_N$. 
We find, however, that with the cuts optimized for Higgs searches very few $N$ events are selected: the cuts used  in
\cite{CMS:2022uhn} keep only 0.7\% of the 9950 events, versus 11\% of the 11320 Higgs events. The basic reason is that the charged lepton from $N\to \mu^+ W^-$ tends to have a $p_{\rm T}$ below the 15 GeV required there to the subleading lepton. If we relax this minimum $p_{\rm T}$ to
5 GeV and impose that the electron is more energetic than the muon we 
would keep 1.1\% of the initial $N$ sample (106 events), while the fraction of $h\to WW^*$ events passing the cuts would be reduced to 4.4\% (500 Higgs events). These estimates suggest a possible complementarity of Higgs physics and $N$ searches in this mass region.

\section{Discussion} 

In any search for a new particle it is key to find the optimal kinematical variable that discriminates between signal 
and background. Here we have discussed how to reconstruct the mass of an HNL in the trilepton plus neutrino channel. This possibility has not been considered in several recent CMS analyses  \cite{CMS:2018iaf,CMS:2024xdq},
where they use a combination of observables optimized with machine learning techniques. Obviously, when the signal events define a mass peak its search is an interesting option.
For example, we may compare our  Fig.~\ref{f3} with Fig.~11 in \cite{CMS:2024xdq}. For $m_N=150$ GeV, the BDT trained with $85$-$150$ GeV masses  gives similar signal and background distributions, in particular, the three bins with the largest fraction of signal collect 53$\%$ of the total, but also 64$\%$ of the background events. For the same HNL mass the BDT trained with $200$-$250$ GeV masses provides a peak-like feature and a better ratio: 27$\%$ of the signal versus just 8$\%$ of the background. On the other hand, the reconstructed mass provides three bins around $m_N$ collecting  $43\%$ of the signal and only $9\%$ of the total background. 
We think that the observable $m_N$ may complement this type of optimized searches at the LHC, in particular, in the mass regions (170 GeV, 270 GeV in \cite{CMS:2024xdq}) of transition between different BDTs.

We have discussed how to solve the two-fold degeneracy in the quadratic equation that provides the longitudinal momentum of the neutrino in the reconstruction of $m_N$. We find that this equation has no solution in a small fraction (around 10\%) of signal events. This is mainly due to the smearing effect introduced by the detector, and also to the different kinematics in signal events where the third lepton comes from $\tau\to \nu\ell\nu$. We find interesting that the background includes a much larger fraction (around 25\%) of events with no solution to the equation, as this could be used as an extra cut also when studying other observables.

In another recent search for a light ($m_N<m_W$) HNL \cite{ATLAS:2022atq} (see also \cite{Dib:2017iva}), ATLAS has reconstructed $m_N$ using that in $pp\to W^+ \to \ell^+ N$ (see Fig.~\ref{f1}) the $W$ is near the 
mass shell. Since the $W$ boson in $N\to \ell^- W^+$ is exactly on shell, our analysis provides a more accurate reconstruction 
in the complementary regime with $m_N>m_W$. We have also shown that a similar reconstruction may work in
the dilepton plus $p_{\rm T}^{\rm miss}$ channel by changing the cuts currently being used in Higgs to $WW^*$ searches. 

In summary, 
the mass peak discussed here could complement and simplify the search for an HNL in $pp$ collisions: even if it is not as {\it clean} as the ones found in Higgs production, its search is also largely  insensitive to normalization uncertainties related to luminosity, overall cross sections, or selection efficiencies. Although it is a possibility not explored at the LHC, our results suggest that it could help to set collider limits on $|V_{\ell N}|^2$ competitive with PMNS unitarity bounds.

\section*{Acknowledgments}
We would like to thank Mikael Chala, David Mu\~noz, Adri\'an Rubio and Jos\'e Santiago for discussions. 
This work has been supported by the Spanish Ministry of Science, Innovation and Universities 
MICIU/AEI/ 10.13039/501100011033/ (grants PID2022-14044NB-C21 and PID2022-139466NB-C22), by Junta de Andaluc{\'\i}a (FQM 101) and by 
Uni\'on Europea-NextGenerationEU (grants $\mathrm{AST22\_6.5}$ and $\mathrm{AST22\_8.4}$).

\end{document}